\documentclass[pra,aps,showpacs,groupedaddress,superscriptaddress,twocolumn,toc=flat,biblatex,footinbib]{revtex4-1}
\usepackage{natbib}
\usepackage[table]{xcolor}
\usepackage[colorlinks=false]{hyperref}
\usepackage[utf8]{inputenc}
\usepackage{color}
\usepackage{bbm} 
\usepackage[english]{babel}
\usepackage{multirow}
\usepackage{amsfonts,amsmath,amssymb,stmaryrd}

\usepackage{graphicx}
\usepackage{subfigure}  

\usepackage{epsfig}
\usepackage{mathrsfs}
\usepackage{verbatim}
\usepackage{centernot}
\usepackage{ulem}
\usepackage{longtable}
\usepackage{nicefrac}
\usepackage[framemethod=tikz]{mdframed}
\usepackage{siunitx}
\usepackage[nolist]{acronym}
 
\usepackage{array}

\usepackage{cancel,ifthen}
\newcommand{\cmnt}[2][NoInPuT]{\ifthenelse{\equal{#1}{NoInPuT}}{}{{\color{red}\sout{#1}}} {\color{blue} #2}}

\usepackage{bm}	
\renewcommand{\vec}[1]{\bm{#1}}
\newcommand{\ak}[1]{{\color{black} #1}}

\LTcapwidth=\textwidth

\begin{document}

\normalem	

\title{Rotationally invariant formulation of spin-lattice coupling in multi-scale modeling}

\author{Markus Weißenhofer}
\email[]{markus.weissenhofer@uni-konstanz.de}
\affiliation{Department of Physics, University of Konstanz, DE-78457 Konstanz, Germany}

\author{Hannah Lange}
\affiliation{Department of Chemistry/Phys. Chemistry, LMU Munich,
Butenandtstrasse 11, D-81377 Munich, Germany}

\author{Akashdeep Kamra}
\affiliation{Condensed Matter Physics Center (IFIMAC) and Departamento de F\'{i}sica Te\'{o}rica de la Materia Condensada, Universidad Aut\'{o}noma de Madrid, E-28049 Madrid, Spain}

\author{Sergiy Mankovsky}
\affiliation{Department of Chemistry/Phys. Chemistry, LMU Munich,
Butenandtstrasse 11, D-81377 Munich, Germany}
\author{Svitlana Polesya}
\affiliation{Department of Chemistry/Phys. Chemistry, LMU Munich,
Butenandtstrasse 11, D-81377 Munich, Germany}
\author{Hubert Ebert}
\affiliation{Department of Chemistry/Phys. Chemistry, LMU Munich,
Butenandtstrasse 11, D-81377 Munich, Germany}

\author{Ulrich Nowak}
\affiliation{Department of Physics, University of Konstanz, DE-78457 Konstanz, Germany}


\date{\today}

\begin{abstract}
In the spirit of multi-scale modeling, we \ak{develop a theoretical framework} for spin-lattice coupling that connects, on the one hand, to ab initio calculations of spin-lattice coupling parameters and, on the other hand, to the magneto-elastic continuum theory. \ak{The derived} Hamiltonian describes a closed system of spin and lattice degrees of freedom \ak{and explicitly conserves} the total momentum, angular momentum and energy. Using a new numerical implementation that corrects earlier Suzuki-Trotter decompositions we perform simulations on the basis of the resulting equations of motion to investigate the combined magnetic and mechanical motion of a ferromagnetic nanoparticle, \ak{thereby validating our developed method}. In addition to the ferromagnetic resonance mode of the spin system we find another low-frequency mechanical response and a rotation of the particle according to the Einstein-de-Haas effect. \ak{The framework developed herein will enable the use of multi-scale modeling for investigating and understanding a broad range of magneto-mechanical phenomena from slow to ultrafast time scales.}
\end{abstract}

\maketitle

\begin{acronym}
\acro{SD}[SD]{spin dynamics}
\acro{SLD}[SLD]{spin-lattice dynamics}
\acro{FMR}[FMR]{ferromagnetic resonance}
\acro{DMI}[DMI]{Dzyaloshinskii-Moriya interaction}
\acro{ME}[ME]{magneto-elastic}
\acro{AM}[angular momentum]{angular momentum}
\acro{SLC}[SLC]{spin-lattice coupling}
\acro{SSC}[SSC]{spin-spin coupling}
\acro{EdH}[EdH]{Einstein-de-Haas}
\end{acronym}


The spin-orbit interaction is a relativistic \ak{effect} \ak{at} the heart of modern spintronics~\cite{Manchon2015}. It couples the electron's spin to its orbital motion and \ak{plays a central role in} quantum materials \ak{bearing high potential} for future nanoelectronic devices. \ak{Its manifestations include phenomena like} magneto-crystalline anisotropy and \ac{DMI}~\cite{Dzyaloshinskii1958,Moriya1960}. While the focus in spintronics has long been on \ak{electrons} or magnons as \ak{carriers} of \acl{AM}, newer lines of research include even circularly polarized phonons to fully understand and control the flow of \acl{AM} in a material \cite{Hirohata2020, Garanin2015, Rueckriegel2020, Mentnik2019}. Recently, it was demonstrated that even on ultrashort time scales \acl{AM} can be transferred from the spin system to the lattice \cite{Tauchert2022}. In the lattice, the spin \acl{AM} is absorbed \ak{by} phonons carrying \acl{AM} till — on larger times scales — the macroscopic \ac{EdH} effect sets in \cite{Dornes2019}. A coupling between spin and lattice degrees of freedom that - beside the exchange of energy - includes the exchange of \acl{AM} must be based on spin-orbit coupling, the effect \ak{of} which has to be taken into account for a complete description of \ac{SLD}.

Descriptions of spintronic phenomena are often based on spin models, which treat the lattice degrees of freedom as a heat bath and define the spin Hamiltonian with its magneto-crystalline anisotropy and DMI for a rigid lattice. Consequently, the spin \acl{AM} is not conserved. First attempts, to develop a framework for the calculation of coupled \ac{SLD} — also referred to as molecular and spin dynamics simulations — suffer from an \ak{incomplete formulation of the spin lattice interaction} ~\cite{Ma2009, Perera2016, Strungaru2021, Assmann2019}. 
The works by Aßmann \cite{Assmann2019} and Strungaru \cite{Strungaru2021} assume a pseudo dipolar coupling that conserves the total \acl{AM} - a prerequisite for a well-defined \ac{SLC}. However, it suffers from the fact that it is not linked to first principles calculations of \ac{SLC} terms, which always rest on an expansion of the spin Hamiltonian with respect to small distortions of the lattice. 
The works by Hellsvik et al. \cite{Hellsvik2019}, Sadhukan et al. \cite{Sadhukhan2022} and Mankovsky et al. \cite{Mankovsky2022}, use exactly these \ac{SLC} terms, that can be derived from first principles. However, these terms do not conserve the total \acl{AM} since they are not rotationally invariant. \ak{This inconsistency and the need for rotational invariance} has already been pointed out 50 years ago in the context of the \ac{ME} theory \cite{Melcher1970,Melcher1972}, a continuum theory that \ak{approximates} a microscopic spin lattice model Hamiltonian on larger length scales.

In this \ak{Letter}, we \ak{develop a rotationally invariant description of spin lattice interaction for multi-scale modeling relaxing} the assumption of a rigid lattice with fixed orientation. The resulting Hamiltonian is translationally and rotationally invariant, keeping total energy, momentum, and \acl{AM} constant. All terms can be linked to the recently developed ab initio methods that allow for a first principles calculation of model parameters \cite{Hellsvik2019,Sadhukhan2022,Mankovsky2022}, opening perspectives for multi-scale modeling of \ac{SLD}.  We also demonstrate that our spin-lattice Hamiltonian represents the discrete formulation of magneto-elastic theory and we link the microscopic parameters with the  magneto-elastic constants. We show that even terms that - in a spin model - do not include any lattice distortions must transfer \acl{AM} to the lattice. Furthermore, we derive the equations of motion for spin and lattice degrees of freedom and solve them numerically with an appropriate  Suzuki-Trotter decomposition. Finlly, we present first simulations of the precession of a magnetized body and spin dynamics including the resulting response of the lattice.

\begin{figure}
    \centering
    \includegraphics[width=0.4\textwidth]{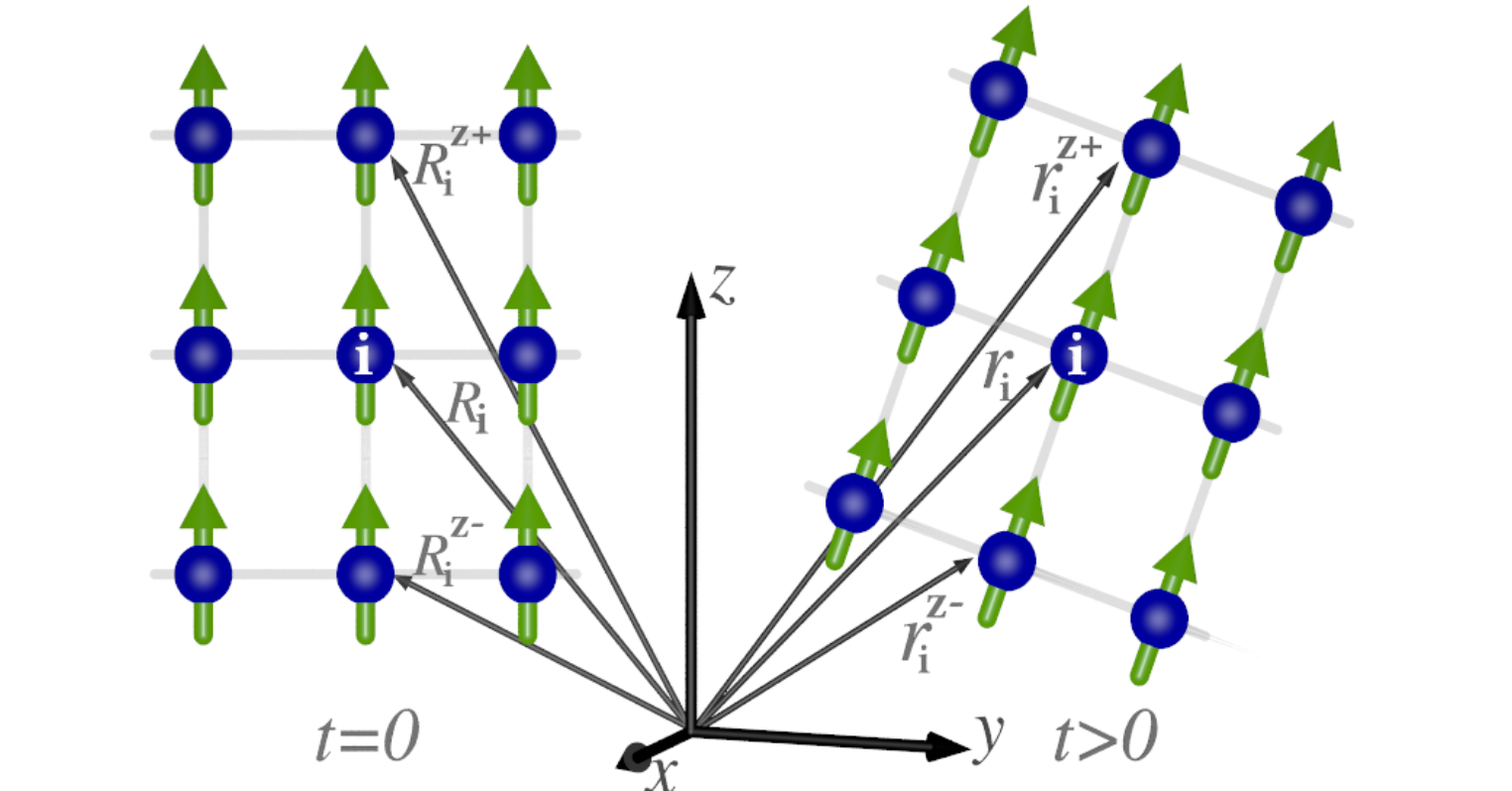}
    \caption[]{Rotation and translation of a magnetized sample. The reference \ak{configuration} at $t=0$ is denoted by $\vec{R}_i$ (left), for $t>0$ by $\vec{r}_i$ (right). During its motion, the easy axis for an atom $i$ at position $\vec{r}_i$ or $\vec{R}_i$ can be defined via its upper and lower neighbors at position ${r}_i^{z\pm}$.}
    \label{fig:e_z_Definition}
\end{figure}

A complete Hamiltonian accounting simultaneously for the spin and lattice degrees of freedom contains contributions from the lattice degrees of freedom (kinetic energy and pair potentials) as well as contributions which include the spin degrees of freedom. The latter can be expressed as an expansion of relativistic spin-spin interactions for small distortions \cite{Hellsvik2019, Mankovsky2022},
    \begin{align} 
    &\mathcal{H}_{\mathrm{SLC}}
    \approx  
    \sum_{ij,\alpha \beta }\mathcal{J}^{\alpha \beta}_{ij}S_i^{\alpha}S_j^{\beta} 
    +  
    \sum_{ijk,\alpha \beta \mu}\mathcal{J}^{\alpha \beta,\mu}_{ij,k}S_i^{\alpha}S_j^{\beta}\left(u_k^{\,\mu}-u_i^{\,\mu}\right)  \nonumber 
    \\
    &+ \sum_{ijkl,\alpha \beta \mu\nu}\mathcal{J}^{\alpha \beta,\mu\nu}_{ij,kl}S_i^{\alpha}S_j^{\beta}\left(u_k^{\,\mu}-u_i^{\,\mu}\right)\left(u_l^{\,\nu}-u_i^{\,\nu}\right),
    \label{eq:H_SLC_1}
\end{align}
where the summation \ak{runs} over the lattice (latin indices) and Cartesian coordinates (greek indices)~\ak{\footnote{In this notation Eq.~\eqref{eq:H_SLC_1} also includes on-site terms (where $i=j$).}}. $\vec{S}_{i}$ are \ak{unit} vectors representing the direction of magnetic moments at sites $i$, and $\vec{u}_i=\vec{r}_i-\vec{R}_i$ are displacement vectors of atoms $i$ at position $\vec{r}_i$ (and equilibrium position $\vec{R}_i$ in a reference \ak{configuration}, see \ak{Fig.~\ref{fig:e_z_Definition}}). The \ac{SSC} $\mathcal{J}^{\alpha \beta}_{ij}$ and \ac{SLC} tensors \ak{ $\mathcal{J}^{\alpha \beta,\mu}_{ij,k}=\partial \mathcal{J}^{\alpha \beta}_{ij}/\partial u_k^{\mu}$} are defined with respect to a chosen coordinate system. As shown by Mankovsky et al., these tensors can be calculated quantitatively from first principles~\cite{Mankovsky2022}. 

The relative displacements $\left(u_k^{\,\mu}-u_i^{\,\mu}\right)$ w.r.t.\ a reference atom $i$ take into account deformations of the lattice. \footnote{Note that in the work of Mankovsky et al.~\cite{Mankovsky2022} these tensors are calculated taking only one displacement at site $k$ into account while all the other atoms are in their equilibrium position. In this case the relative displacement is $u_k^{\,\mu} - u_i^{\,\mu} = u_k^{\,\mu}$.  However, for a system that is displaced as a whole (equally for all sites $i$) the relative displacement will vanish and there is no additional contribution to the potential energy.}
They are the discrete lattice representation of the strain and rotation tensor elements of elasticity theory. As such, Eq.~\eqref{eq:H_SLC_1} represents the discrete formulation of \ak{the} \ac{ME} theory \cite{Kittel1949}, from which we can derive an extended expression for the \ac{ME} energy density,
\begin{align}
    \epsilon
    &=
    \sum_{\alpha\beta\mu\nu}
    S^\alpha
    S^\beta
    \Big(
    {B}^\mathrm{s}_{\alpha\beta,\mu\nu}
    \varepsilon_{\mu\nu}
    +
    {B}^\mathrm{as}_{\alpha\beta,\mu\nu}
    \omega_{\mu\nu}
    \Big) \nonumber
    \\
    &+
   \sum_{\alpha\beta\gamma\mu\nu}
   \partial_\beta S^\alpha
   \partial_\gamma S^\alpha
   \Big(
   A^\mathrm{s}_{\beta\gamma,\mu\nu}
   \varepsilon_{\mu\nu}
   +
   A^\mathrm{as}_{\beta\gamma,\mu\nu}
   \omega_{\mu\nu}
   \Big)
   \label{eq:me_energy}
   \\
   &+
    \sum_{\alpha\beta\gamma\delta\mu\nu}
    \varepsilon^{\alpha\beta\gamma}
    S^\alpha
    \partial_\delta S^\beta
    \Big(
    D^\mathrm{s}_{\gamma\delta,\mu\nu}
    \varepsilon_{\mu\nu}
    +
    D^\mathrm{as}_{\gamma\delta,\mu\nu}
    \omega_{\mu\nu}
    \Big),\nonumber
\end{align}
where $\vec{S}$ is the continuous magnetization, $\varepsilon_{\mu\nu}$ the strain tensor, $\varepsilon^{\alpha\beta\gamma}$ the Levi-Civita symbol, and $\omega_{\mu\nu}$ the rotation tensor. The \ak{important role of} the latter \ak{in the} \ac{ME} theory has been addressed before by Melcher~\cite{Melcher1970,Melcher1972}, \ak{and reaffirmed in recent experiments~\cite{Xu2020,Kuess2020,Kuess2022}}. The terms in Eq.~\eqref{eq:me_energy} model anisotropy, Heisenberg exchange, and \ac{DMI} due to lattice distortions and \ak{the} corresponding symmetric/antisymmetric \ac{ME} tensors ${B}^\mathrm{s/as}_{\alpha\beta,\mu\nu}$, $ A^\mathrm{s/as}_{\beta\gamma,\mu\nu}$, and $D^\mathrm{s/as}_{\gamma\delta,\mu\nu}$ can be obtained from the microscopic SLC tensors. \ak{A detailed} derivation can be found in the Supplemental Material~\cite{SupplMat} as well as the connection between the \ac{ME} constants \cite{Kittel1949} and the microscopic SLC tensors.


Looking at Eq.~\eqref{eq:H_SLC_1} one finds immediately that this Hamiltonian does not conserve the total (spin and lattice) \acl{AM}, since it is not rotationally invariant. It is, hence, not capable of describing spins \ak{plus lattice as a} closed system. \ak{To understand this, we examine} an isotropic Heisenberg model with a uniaxial on-site anisotropy,
\begin{align}
    \mathcal{H}_{ani} = -\sum_{ij} J_{ij} \vec{S}_i \cdot  \vec{S}_j - d_z\sum_i \left(S_i^z\right)^2,
    \label{eq:H_ani_1}
\end{align}
for a system with the $z$ axis being the easy axis of the magnetization. \ak{Here,} the Heisenberg exchange interaction term is rotationally invariant and conserves the total spin \acl{AM}. The anisotropy term, however, is not rotationally invariant and the total spin \acl{AM} is, hence, not conserved. To keep the total \acl{AM} conserved, the spin \acl{AM} would have to go to the lattice but since this term does not include any lattice degrees of freedom it cannot. 

The situation is shown in Fig.~\ref{fig:e_z_Definition}. Let us assume the atoms are at time $t=0$ in equilibrium positions $\vec{R}_i$ in a reference \ak{configuration} with the easy axis along $z$ \ak{(defined in the lab frame)}. When the sample starts moving the lattice the atom positions at later times are $\vec{r}_i(t) = \vec{R}_i +\vec{u}_i(t)$ and the easy axis may no longer be aligned with the $z$-axis of the \ak{lab} frame. Consequently, the anisotropy term in Eq.~\eqref{eq:H_ani_1} has to be transformed. This can be done by projecting the spin orientation using unit vectors that are \ak{defined in terms of the respective neighbor atoms}.

In a cubic lattice an appropriate unit vector $\vec{e}^{z}$ can be defined via the upper ($+$) and lower ($-$) nearest neighbors of atom $i$ at positions $\vec{r}_i^{z \pm}=\vec{R_i}^{z \pm}+\vec{u_i}^{z\pm}$, yielding
\begin{equation}
    \vec{e}_i^{z(\pm)} = 
    \frac{\vec{r_i}^{z\pm} - \vec{r}_i}{\vert \vec{r}_i^{z \pm} - \vec{r}_i \vert }\, .
    \label{eq:e_zplus}
\end{equation}
Now we can \ak{write} the anisotropy term of Eq.~\eqref{eq:H_ani_1} \ak{without reference to a specific coordinate system} and we obtain  
\begin{align}
    \mathcal{H}_{ani}&=-\sum_{ij} J_{ij} \vec{S}_i \cdot  \vec{S}_j  -\frac{d_z}{2}\sum_i \Big[\Big(\vec{S}_i\cdot \frac{\vec{r}_i^{z+}-\vec{r}_i}{\vert\vec{r}_i^{z+}-\vec{r}_i \vert }\Big)^2
    \nonumber \\
    & +\Big(\vec{S}_i\cdot \frac{\vec{r}_i^{z-}-\vec{r}_i}{\vert\vec{r}_i^{z-}-\vec{r}_i \vert }\Big)^2 \Big].
    \label{eq:H_ani_2}
\end{align}
The resulting Hamiltonian contains only scalar products of the spins $\vec{S}_i$ and differences of position vectors $\vec{r}_i$. It is hence translationally and rotationally invariant and will keep the total momentum and \acl{AM} constant. Most \ak{importantly}, the transformed Hamiltonian contains the lattice degrees of freedom \ak{explicitly}, even though the original one did not. Only this makes it possible to transfer \acl{AM} from the spins to the lattice, keeping the total \acl{AM} constant. 

The \ak{microscopic} origin of the \ak{anisotropy} can be crucial in determining the exact form of the definition of the $z$ direction. It is however sufficient to take only two neighbors into account to capture the majority of effects. Note that this local definition of an easy axis does not only work for a global rotation but also for deformations of the sample. Furthermore, both, upper and lower neighbors, are used to define the local easy axis for spins $i$, a definition that \ak{holds} also at surfaces with a reduced number of neighbors. Nevertheless, it should be stressed that this definition is neither unique nor trivial, since the choice of neighbors for the definition of the unit vectors will affect the equations of motion and \ak{the} atoms the \acl{AM} is transferred to.

The transformation above can be extended to other contributions of the spin-lattice Hamiltonian \eqref{eq:H_SLC_1}. In order to do so, the local definition for the unit vector in $z$ direction for atom $i$ from Eqs. \eqref{eq:e_zplus} can be generalized to a set of three orthogonal directions $\alpha$,
\begin{align}
    \vec{e}_i^{\alpha(\pm)}=\frac{\vec{r_i}^{\alpha \pm}-\vec{r}_i^{\alpha}}{\vert\vec{r}_i^{\alpha \pm}-\vec{r}_i^{\alpha} \vert }.
    \label{eq:e_alpha}
\end{align}
Similar to the case of a uniaxial anisotropy, these unit vectors can be used to transform the first term of Eq.~\eqref{eq:H_SLC_1},
\begin{align}
    \mathcal{H}_{\mathrm{SS}}&=\sum_{ij}\sum_{\alpha \beta }\mathcal{J}^{\alpha \beta}_{ij}S_i^{\alpha}S_j^{\beta}=\sum_{ij}\sum_{\alpha \beta }\mathcal{J}^{\alpha \beta}_{ij}(\vec{S}_i\cdot \vec{e}_i^{\,\alpha})(\vec{S}_j\cdot \vec{e}_j^{\,\beta}) 
    \label{eq:H_Heisenberg}
\end{align}
Again, this Hamiltonian consists of \ak{scalar} products of \ak{spins} and differences of position vectors and is hence rotationally invariant. Analogously, the spin-lattice Hamiltonian becomes
\begin{align}
\mathcal{H}_{\mathrm{SLC}}
&= 
\mathcal{H}_{\mathrm{SS}}
+ 
\sum_{ijk,\alpha \beta \mu}\mathcal{J}^{\alpha \beta,\mu}_{ij,k}(\vec{S}_i\cdot \vec{e}_i^{\,\alpha})(\vec{S}_j\cdot \vec{e}_j^{\,\beta}) \times
\nonumber
\\ & \times \big[\left(\vec{r}_k-\vec{r}_i\right)\cdot \vec{e}_k^{\,\mu} - R_{ki}\big]+\dots \,,
    \label{eq:H_SLC_3}    
\end{align}
where $R_{ki}$ is the equilibrium distance between atoms $k$ and $i$ in the reference \ak{configuration}. This Hamiltonian consists of the rotationally invariant spin-spin term ($\mathcal{H}_{\mathrm{SS}}$) and a second spin-spin-lattice term.

Each term in the initial formulation \eqref{eq:H_SLC_1} that breaks rotational symmetry now depends on the spins and the lattice positions and, hence, can transfer \acl{AM} between the two subsystems. Thus, the dominating terms for \acl{AM} transfer may vary for different materials, depending on the specific values of the \ac{SSC} and \ac{SLC} tensors. \ak{For example,} in Fe the transfer is mainly via the spin-lattice \ac{DMI} \cite{Mankovsky2022}, whereas in FePt the dominating terms are two-site anisotropy terms (see Table~\ref{tab:SSC_SLC}).

\begin{table}[]
\begin{tabular}{c|c|cccc}
   \ak{Material} & \ak{Contribution} to \ac{SSC} & $\vert J_{ij}^{\mathrm{iso}}\vert $      &   $\vert J_{ij}^{\mathrm{dia-a}}\vert $    & $\vert J_{ij}^{\mathrm{off-s}}\vert $      &   $\vert \vec{D}_{ij}\vert$    \\\hline
   \\[-1em]
\multirow{2}{*}{Fe} & $J_{ij}^{SSC}$ & 11.389 & 0.019 & 0.017 & 0.0 \\
& $\Delta J_{ij}^{SLC}(u_j^x)$ & 1.587 & 0.002 & 0.003 & 0.062 
\\[-1em]
\\\hline
\\[-1em]
\multirow{2}{*}{FePt} &$J_{ij}^{SSC}$ & 9.590 & 0.320 & 0.209 & 0.0 \\
&$\Delta J_{ij}^{SLC}(u_j^x)$ & 1.960 & 0.023 & 0.024 & 0.089
\\[-1em]
\end{tabular}
\vspace{0.5em}
\caption{Maximal absolute \ac{SSC} $J_{ij}^{SSC}$ and maximal modification of \ac{SSC} due to \ac{SLC} $\Delta J_{ij}^{SLC}=J_{ij,j}\cdot u_j^x$ (in meV) in the presence of a displacement $u_j^x=0.03\,a_{\mathrm{lat}}$ in Fe and FePt for different \ac{SSC} contributions. For both materials, we consider sites $i$ and $j$ being Fe atoms with different distances $r_{ij}$ and list the values for the pair $ij$ with the largest contribution to the respective parts of the \ac{SSC} tensor. In Fe, the largest contribution which can transfer \acl{AM} is the spin-lattice \ac{DMI} $ \vert \Delta \vec{D}_{ij}\vert= \vert \vec{D}_{ij,j}^x \cdot u_j^x\vert $ for $r_{ij}=1\,a_{\mathrm{lat}}$, in FePt it is the spin-spin anti-symmetric diagonal part $\vert J_{ij}^{\mathrm{dia-a}}\vert =\frac{1}{2}\vert  J_{ij}^{xx}-J_{ij}^{zz}\vert$ for $r_{ij}=1.414\,a_{\mathrm{lat}}$. The symmetrized off-diagonal elements are defined as $J_{ij}^{\mathrm{off-s}}=\frac{1}{2}\left(J_{ij}^{xy}+J_{ij}^{yx}\right)$. For details see the Supplemental \ak{Material~\cite{SupplMat}.}}
\label{tab:SSC_SLC}
\end{table}

As a first application of our formulation and to test its validity, we perform combined \ac{SLD} simulations using the following Hamiltonian for a simple cubic lattice,
\begin{align}
        \mathcal{H}=&\mathcal{H}_{ani} 
        + \sum_{i} \frac{\vec{p}_i^2}{2 m}
        + V_0 \sum_{ij} \frac{(r_{ij}-R_{ij})^2}{R_{ij}} 
    \label{eq:Hamiltonian_simulations}
\end{align}
that extends our rotationally invariant formulation of the spin Hamiltonian of Eq. \eqref{eq:H_ani_2} by terms describing the interaction and the kinetic energy of the lattice, with $m$ being the mass of the atoms and $V_0$ describing the strength of the lattice interactions in \ak{the} harmonic approximation. For the sake of simplicity, we assume that these interactions are restricted to the first three shells of neighbors and that they scale inversely with the equilibrium distance.

Evaluating the dynamics of spin and lattice degrees of freedom $\{ \vec{r}_i,\vec{p}_i,\vec{S}_i \}$ requires the concurrent solution of the coupled equations of motion,
\begin{align}
        \dot{\vec{r}}_i = \frac{\partial \mathcal{H}}{\partial \vec{p}_i},
        \hspace*{0.66em}
        \dot{\vec{p}}_i = -\frac{\partial \mathcal{H}}{\partial \vec{r_i}},
        \hspace*{0.66em}
        \mathrm{and}
        \hspace*{0.66em}
        \dot{\vec{S}}_i=\frac{\gamma}{\mu_\mathrm{s}} \vec{S}_i \times \frac{\partial \mathcal{H}}{\partial \vec{S}_i},   
\end{align}
with $\gamma$ and $\mu_\mathrm{s}$ being the absolute values of the gyromagnetic ratio and the magnetic spin moment, respectively. Conservation of energy, momentum and \acl{AM} can be ensured by using a symplectic algorithm. Here, we use a scheme based on the Liouville formalism \cite{Frenkel2001} and the Suzuki-Trotter decomposition \cite{Suzuki1976} that was initially proposed in Ref.~\cite{Omelyan2001} and has proven reliable for the simulation of combined \ac{SLD} \cite{Tsai2005,Ma2009,Ma2016,Assmann2019,Strungaru2021}. Note that the presence of a uniaxial on-site anisotropy term, which is quadratic in the spins, requires a further decomposition of the integration scheme that has not been discussed in literature so far. Details, tests of the conservation of the total \acl{AM} and the energy of the system and a comparison of the temperature dependence of the magnetization with \ac{SD} simulations based on the stochastic Landau-Lifshitz-Gilbert equation of motion can be found in the Supplemental \ak{Material~\cite{SupplMat}}.

\begin{figure}[t]
    \centering
\includegraphics[width=0.48\textwidth]{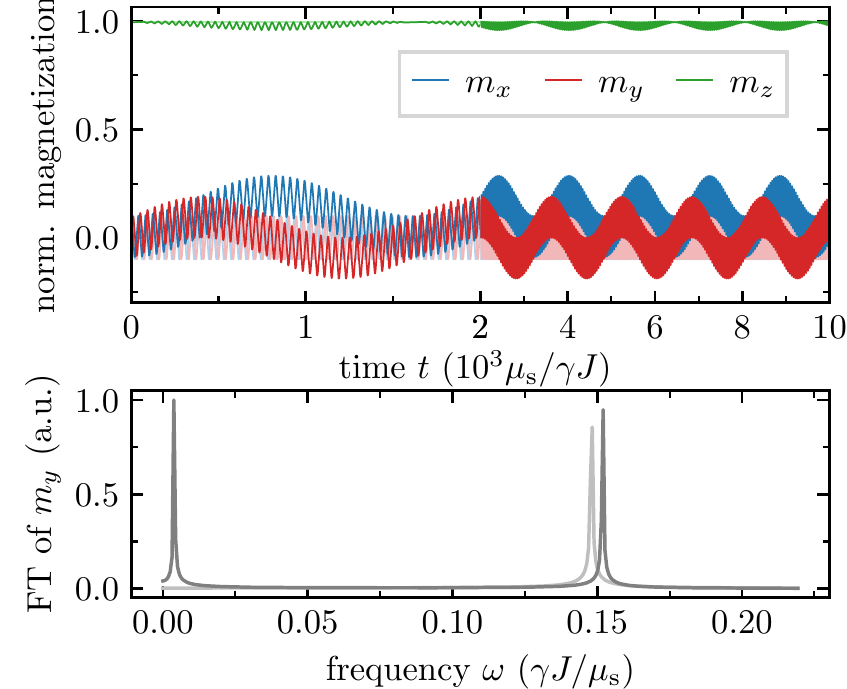}
    \caption{Coherent magnetization dynamics of a free cubic nanoparticle obtained from \ac{SLD} in comparison with \ac{SD} simulations (light curves). Top: magnetization vector components $m_\alpha$ versus time. Bottom: Fourier transform of $m_y$.}
    \label{fig:coherent_spin_rotation}
\end{figure}

As a first application we study the coupled magnetization and lattice dynamics of a free cubic nanoparticle. For this simulation we assume that initially the cube is oriented such that the easy axis is aligned with the $z$-axis and all spins point along $\vec{m}_0 = (0.1,0,\sqrt{1-0.1^2})^\mathrm{T}$. This gives rise to a coherent precession of the magnetization along with mechanical motion of the cube. Fig.~\ref{fig:coherent_spin_rotation} displays the time evolution of the magnetization $\vec{m} = \frac{1}{N}\sum_i^N \vec{S}_i$ and the Fourier transform of its $y$-component for a nanoparticle consisting of $4^3$ atoms. The light curves are obtained by pure \ac{SD} simulations, for which we keep the position of the atoms \ak{fixed}. In contrast to pure \ac{SD}, the \ac{SLD} simulations produce oscillations at two characteristic frequencies $\omega_{\vec{n}_3}\approx 3.89\times 10^{-3} \gamma J/\mu_\mathrm{s}$ and $\omega_\mathrm{FMR}\approx 0.152 \gamma J/\mu_\mathrm{s}$. The peak at $\omega_\mathrm{FMR}$ can be attributed to the usual \ac{FMR} frequency and is close to the value predicted by linear spin wave theory $\omega_\mathrm{FMR}=2\Bar{d}\gamma/\mu_\mathrm{s}=0.15 \gamma J/\mu_\mathrm{s}$,  where $\Bar{d}$ is the averaged uniaxial magnetic \ak{anisotropy~\footnote{Given the form of the anisotropy in Eq.~\eqref{eq:H_ani_2}, the effective uniaxial anisotropy of spins at two faces of the cube is reduced by a factor of two. Thus, for a cube consisting of $4^3$ atoms, we calculate $\Bar{d}=0.75 d_z$.}}. The \ac{SD} value ($\omega_\mathrm{FMR}\approx 0.148 \gamma J/\mu_\mathrm{s}$) is slightly smaller due to finite size effects: the spins at the edges of the cube lag behind, since their anisotropy field is weaker [cf.\ Eq.~\eqref{eq:H_ani_2}], slowing down the overall precession frequency. Surprisingly, the \ac{FMR} frequency of \ac{SLD} is shifted to higher values as compared to the \ac{SD} value.

\begin{figure}
    \centering
      \includegraphics[width=0.48\textwidth]{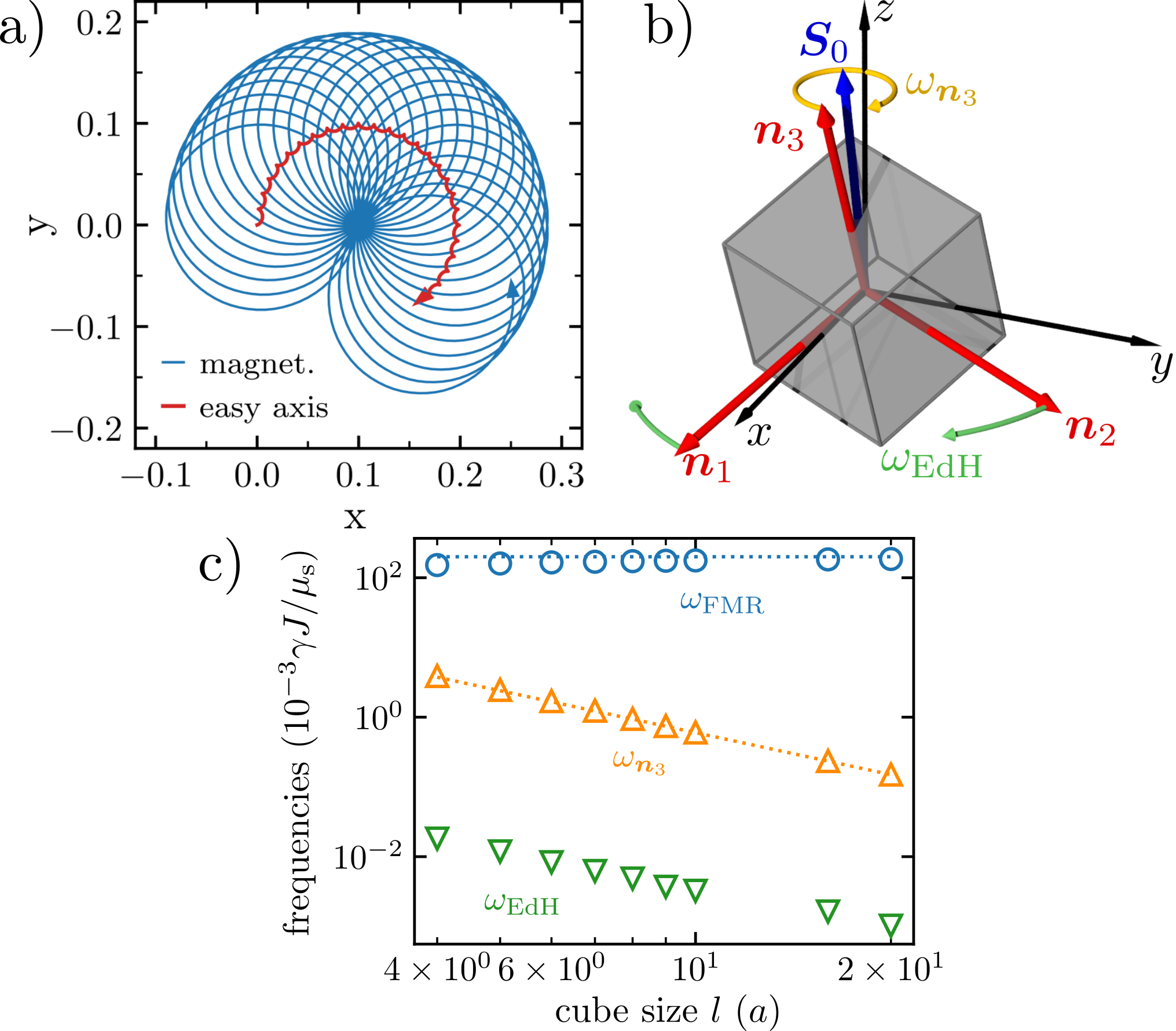}
    \caption{Mechanical motion of a cubic nanoparticle excited by coherent precession of the magnetization around $\vec{n}_3$ (easy axis). a) Spiraling magnetization dynamics and easy axis precession in the time interval $[0,10^3]\mu_\mathrm{s}/\gamma J$. b) Sketch of the two characteristic mechanical modes. c) Characteristic frequencies versus cube size $l$.  Dotted lines correspond to theory curves as explained in the text.}
    \label{fig:mechanical_motion}
\end{figure}

The emergence of the peak at $\omega_{\vec{n}_3}$ as well as the shift of the \ac{FMR} frequency are a result of the mechanical motion of the nanoparticle, which can be characterized \ak{by} the vectors $\vec{n}_\alpha$ \ak{($\alpha\in{1,2,3}$) normal} to the faces of the cube. \ak{As} the easy axis is firmly attached to one of these vectors (without loss of generality, $\vec{n}_3$), we can compare the dynamics of the easy axis and the magnetization, see Fig.~\ref{fig:mechanical_motion}a. We find that the magnetization precesses around the easy axis, which itself revolves around $\vec{m}_0$ exactly at $\omega_{\vec{n}_3}$, giving rise to the second peak in Fig.~\ref{fig:coherent_spin_rotation} and the shift of the \ac{FMR} frequency.

\ak{This} emergence of the easy-axis-precession was predicted in Ref.~\cite{Usov2015} based on a simple rigid-body-macrospin model \cite{Usov2012,Usadel2015,Keshtgar2017}. There, the magnetic nanoparticle is described by the normalized magnetization $\vec{m}$ and the vectors $\vec{n}_\alpha$ introduced above. The dynamics of these vectors $\vec{n}_\alpha$ are given by $\dot{\vec{n}}_\alpha=\vec{\omega} \times \vec{n}_\alpha$, where $\vec{\omega}$ is the angular velocity of the nanoparticle in the laboratory frame, which is related to its \acl{AM} via $\vec{L}=\Theta \vec{\omega}$. For a cubic nanoparticle, the moment of inertia is given by $\Theta=\frac{1}{6} N m l^2$, $l$ being the cube size and $N$ being the number of atoms. Conservation of \acl{AM} requires that $-N \frac{\mu_\mathrm{s}}{\gamma} \dot{\vec{m}} + \dot{\vec{L}} = 0$. If the nanoparticle is initially at rest, we get $\vec{\omega}(t)=\frac{6 \mu_\mathrm{s}}{m l^2\gamma}(\vec{m}(t)-\vec{m}_0)$, with $\vec{m}_0$ being the initial orientation of the magnetization. This yields $\dot{\vec{n}}_\alpha=\frac{6 \mu_\mathrm{s}}{ml^2\gamma}(\vec{m}(t)\times \vec{n}_\alpha-\vec{m}_0\times \vec{n}_\alpha)$.
The first contribution to the torque depends on the current value of the magnetization and amounts to zero for $\vec{n}_3$, due to the rapid oscillations of $\vec{m}(t)$ around $\vec{n}_3$. The second term describes a simple rotation of $\vec{n}_3$ around the initial direction of the magnetization $\vec{m}_0$ with frequency $\omega_{\vec{n}_3}=\frac{6 \mu_\mathrm{s}}{ml^2\gamma}$. For the parameters used here, we calculate $\omega_{\vec{n}_3} \approx 3.75\times 10^{-3} \gamma J/\mu_\mathrm{s}$, which is in close agreement with the simulation results.

In addition to the precession of $\vec{n}_3$, we find an \ac{EdH}-type rotation of $\vec{n}_{1,2}$ with $\omega^\mathrm{EdH} \approx 1.81\times 10^{-5} \gamma J/\mu_\mathrm{s}$ around the $z$-axis of the laboratory frame (see Fig.~\ref{fig:mechanical_motion}b). This rotation occurs when the average magnetization differs from its initial value, since the average angular velocity of the cubic nanoparticle is given by $\langle \vec{\omega}(t)\rangle  =\frac{6 \mu_\mathrm{s}}{ml^2\gamma} \langle (\vec{m}(t)-\vec{m}_0)\rangle$. This leads to a nonzero value only for the $z$-component of the angular velocity (cf.\ Fig.~\ref{fig:coherent_spin_rotation}). 

Fig.~\ref{fig:mechanical_motion}c displays all three characteristic frequencies versus cube size. The \ac{FMR} frequency is the highest and approaches the bulk value for large cubes. The other two (mechanical) frequencies scale with $l^{-2}$ and have a constant ratio over the range considered here. \ak{Besides testing} the validity of the analytical expression for $\omega_{\vec{n}_3}$, this allows us to estimate the mechanical frequencies for real materials. E.g., for an FePt nano-cube with edge length of $\SI{100}{\nano\metre}$ we get  $\omega_{\vec{n}_3} \approx \SI{250}{\kilo\hertz}$ and $\omega_{\mathrm{EdH}} \approx \SI{1}{\kilo\hertz}$ using $\mu_\mathrm{s}\approx 3.23 \mu_\mathrm{B}$, $\gamma=\SI{1.76e11}{\second^{-1}\tesla^{-1}}$ and $m=m_\mathrm{Fe}+m_\mathrm{Pt}\approx \SI{4.17e-25}{\kilogram}$ (the \ac{FMR} frequency for a highly coercive material such as FePt is of the order of $\SI{100}{\giga\hertz}$) \cite{Barker2010}. 

In summary, we have developed a rotationally invariant formulation of coupled spin-lattice dynamics for multi-scale modeling of magneto-mechanical motion. It successfully integrates first principles evaluation of \ac{SLC} parameters, \ac{ME} continuum theory, and spin lattice dynamics simulations. Employing our developed framework and a new numerical implementation that corrects earlier Suzuki-Trotter decompositions we simulate combined magneto-mechanical dynamics of a ferromagnetic nanoparticle, thereby validating our formulation. Our simulations demonstrate that in addition to the ferromagnetic resonance mode of the spin system there are two low-frequency mechanical modes describing the precession of the easy axis and a rotation of the particle according to the \ac{EdH} effect. By incorporating total \acl{AM} conservation, our work provides the tools for simulation of a broad range of magneto-mechanical phenomena. Therefore it is crucial to the understanding of recent and ongoing intriguing experiments, e.g.\ on magnon-phonon coupling or ultrafast magnetization dynamics.

\emph{Acknowledgements.--}
Work in Konstanz is supported by the Deutsche Forschungsgemeinschaft (DFG) via SFB 1432 and Project No. NO 290/5-1. A.K. acknowledges financial support from the Spanish Ministry for Science and Innovation -- AEI Grant CEX2018-000805-M (through the ``Maria de Maeztu'' Programme for Units of Excellence in R\&D). Work in Munich is supported by the DFG via SFB 1277.

%

\end{document}


\begin{acronym}
\acro{SD}[SD]{spin dynamics}
\acro{SLD}[SLD]{spin lattice dynamics}
\acro{FMR}[FMR]{ferromagnetic resonance}
\acro{DMI}[DMI]{Dzyaloshinskii-Moriya interaction}
\acro{ME}[ME]{magneto-elastic}
\end{acronym}

\ak{

\title{Supplemental Material: Rotationally invariant formulation of spin-lattice coupling in multi-scale modeling}

\author{Markus Weißenhofer}
\affiliation{Department of Physics, University of Konstanz, DE-78457 Konstanz, Germany}

\author{Hannah Lange}
\affiliation{Department of Chemistry/Phys. Chemistry, LMU Munich,
Butenandtstrasse 11, D-81377 Munich, Germany}

\author{Akashdeep Kamra}
\affiliation{Condensed Matter Physics Center (IFIMAC) and Departamento de F\'{i}sica Te\'{o}rica de la Materia Condensada, Universidad Aut\'{o}noma de Madrid, E-28049 Madrid, Spain}

\author{Sergiy Mankovsky}
\affiliation{Department of Chemistry/Phys. Chemistry, LMU Munich,
Butenandtstrasse 11, D-81377 Munich, Germany}
\author{Svitlana Polesya}
\affiliation{Department of Chemistry/Phys. Chemistry, LMU Munich,
Butenandtstrasse 11, D-81377 Munich, Germany}
\author{Hubert Ebert}
\affiliation{Department of Chemistry/Phys. Chemistry, LMU Munich,
Butenandtstrasse 11, D-81377 Munich, Germany}

\author{Ulrich Nowak}
\affiliation{Department of Physics, University of Konstanz, DE-78457 Konstanz, Germany}

\maketitle
}

\section{Connection to magneto-elastic theory}
In a phenomenological description the spin-lattice interactions are modeled by the \ac{ME} energy density \cite{Kittel1949,LeCraw1965,Gurevich2020},
\begin{align}
    \begin{split}
    \mathcal{E}_\mathrm{me}
    &=
    B_1 
    \sum_{\alpha}
    S^\alpha
    S^\alpha
    \varepsilon_{\alpha\alpha}
    +
    B_2
    \sum_{\alpha}
    \sum_{\beta, \beta\neq\alpha}
    S^\alpha
    S^\beta
    \varepsilon_{\alpha\beta}
    +
    A_1
    \sum_{\alpha\beta}
    \sum_{\mu,\mu\neq\beta}
    \partial_\beta S^\alpha
    \partial_\mu S^\alpha
    \varepsilon_{\beta\mu}
    +
    A_2
    \sum_{\alpha\beta}
    (\partial_\beta S^\alpha)^2
    \varepsilon_{\beta\beta},
     \label{eq:me_energy_SM}
    \end{split}
\end{align}
with the contiuous magnetization $\vec{S}$ and the lattice deformations in terms of the symmetric strain tensor $\varepsilon$. The parameters $B_1$, $B_2$, $A_1$ and $A_2$ are called \ac{ME} constants. The former two are related to anisotropy effects and the latter two result from the exchange interaction.

Hereinafter, we relate the spin-lattice Hamiltonian introduced in Refs.~\cite{Hellsvik2019,Mankovsky2022} to the phenomenological \ac{ME} theory and derive closed expressions for the calculation of the \ac{ME} constants. We start with the first order correction term to the spin-spin interactions,
\begin{align}
    \mathcal{H}_\mathrm{ssl}
    &=
    \sum_{ijk}
    \sum_{\alpha\beta\mu}
    \mathcal{J}_{ij,k}^{\alpha \beta, \mu}
    S_i^\alpha
    S_j^\beta
    \Big(u_k^\mu-u_i^\mu\Big).
    \label{eq:ssl_hamiltonian_SM}
\end{align}
If spatial variations of the displacements are low, we can approximate $u_k^\mu\approx u_i^\mu+\sum_\nu R^\nu_{ik}\partial_\nu u_i^\mu$, with $\vec{R}_{ik}$ being the vector connecting the equilibrium positions of atoms $i$ and $k$ in the reference frame and $\partial_\nu= \partial/\partial x_\nu$ being the gradient. Eq.~\eqref{eq:ssl_hamiltonian_SM} then becomes
\begin{align}
    \mathcal{H}_\mathrm{ssl}
    &\approx
    \sum_{ijk}
    \sum_{\alpha\beta\mu\nu}
    \mathcal{J}_{ij,k}^{\alpha \beta, \mu}
    S_i^\alpha
    S_j^\beta
    R^\nu_{ik}\partial_\nu u_i^\mu.
    \label{eq:expanded_ssl_SM}
\end{align}
The only terms contributing to $B_1$ and $B_2$ are those proportional to $\mathcal{J}_{ii,k}^{\alpha \beta, \mu}$, i.e.\ the terms that describe the induced anisotropy at site $i$ due to a displacement at site $k$. Going to the continuum limit, we obtain
\begin{align}
    \mathcal{H}_\mathrm{ssl}^\mathrm{ani}
    &=
    \sum_{i}
    \sum_{\alpha\beta\mu\nu}
    S_i^\alpha
    S_i^\beta
    \sum_{k\in \mathrm{N}(i)}
    \mathcal{J}_{ii,k}^{\alpha \beta, \mu}
    \Big(R^\nu_{ik}\partial_\nu u_i^\mu\Big)
    \approx
    \frac{1}{V_c}
    \int \mathrm{d}V\
    \sum_{\alpha\beta\mu\nu}
    S^\alpha(\vec{r})
    S^\beta(\vec{r})
    \partial_\nu u^\mu(\vec{r})
    \sum_{k\in \mathrm{N}(i)}
    \mathcal{J}_{ii,k}^{\alpha \beta, \mu}
    R^\nu_{ik}
    \\
    &=
    \int \mathrm{d}V\
    \sum_{\alpha\beta\mu\nu}
    S^\alpha(\vec{r})
    S^\beta(\vec{r})
    \partial_\nu u^\mu(\vec{r})
   {B}_{\alpha\beta,\mu\nu},
\end{align}
which describes an integration over an energy density $\mathcal{E}^\mathrm{ani}$.

In the above calculations, we introduced the \ac{ME} anisotropy tensor
\begin{equation}
    {B}_{\alpha\beta,\mu\nu}
    =
    \frac{1}{V_c}
     \sum_{k\in \mathrm{N}(i)}
    \mathcal{J}_{ii,k}^{\alpha \beta, \mu}
    R^\nu_{ik},
\end{equation}
with $V_c$ being the volume of the unit cell and where the summation is over the neighbors of atom $i$. Note that $i$ is arbitrary, due to the translational symmetry of the lattice.
The \ac{ME} anisotropy tensor can be separated into a spatially symmetric and an antisymmetric part via

\begin{align}
    {B}^\mathrm{s}_{\alpha\beta,\mu\nu}
    & =
    \frac{{B}_{\alpha\beta,\mu\nu}+{B}_{\alpha\beta,\nu\mu}}{2}
    =
    \frac{1}{2V_c}
     \sum_{k\in \mathrm{N}(i)}
     (
    \mathcal{J}_{ii,k}^{\alpha \beta, \mu}
    R^\nu_{ik}
    +
    \mathcal{J}_{ii,k}^{\alpha \beta, \nu}
    R^\mu_{ik}
    )
    \\
    {B}^\mathrm{as}_{\alpha\beta,\mu\nu}
    &=
    \frac{{B}_{\alpha\beta,\mu\nu}-{B}_{\alpha\beta,\nu\mu}}{2}
    =\frac{1}{2V_c}
     \sum_{k\in \mathrm{N}(i)}
     (
    \mathcal{J}_{ii,k}^{\alpha \beta, \mu}
    R^\nu_{ik}
    -
    \mathcal{J}_{ii,k}^{\alpha \beta, \nu}
    R^\mu_{ik}
    ).
\end{align}
The \ac{ME} energy density associated with anisotropy henceforth becomes
\begin{align}
    \epsilon^\mathrm{ani}
    &=
     \sum_{\alpha\beta\mu\nu}
    S^\alpha
    S^\beta
    {B}_{\alpha\beta,\mu\nu}
    \partial_\nu u^\mu
    =
    \sum_{\alpha\beta\mu\nu}
    S^\alpha
    S^\beta
    \Big(
    {B}^\mathrm{s}_{\alpha\beta,\mu\nu}
    +
    {B}^\mathrm{as}_{\alpha\beta,\mu\nu}
    \Big)
    \partial_\nu u^\mu
    \\
    &=
    \sum_{\alpha\beta\mu\nu}
    S^\alpha
    S^\beta
    \Big(
    {B}^\mathrm{s}_{\alpha\beta,\mu\nu}
    \frac{\partial_\nu u^\mu+ \partial_\mu u^\nu}{2}
    +
    {B}^\mathrm{as}_{\alpha\beta,\mu\nu}
    \frac{\partial_\nu u^\mu- \partial_\mu u^\nu}{2}
    \Big)
    =
    \sum_{\alpha\beta\mu\nu}
    S^\alpha
    S^\beta
    \Big(
    {B}^\mathrm{s}_{\alpha\beta,\mu\nu}
    \varepsilon_{\mu\nu}
    +
    {B}^\mathrm{as}_{\alpha\beta,\mu\nu}
    \omega_{\mu\nu}
    \Big)
\end{align}
where $\varepsilon_{\mu\nu}=\frac{1}{2}(\partial_\nu u^\mu+ \partial_\mu u^\nu)$ and $\omega_{\mu\nu}=\frac{1}{2}(\partial_\nu u^\mu- \partial_\mu u^\nu)$ are the infinitesimal strain and rotation tensors, respectively. Note that we use $\epsilon$ to denote energy densities that are derived from the microscopic Hamiltonian and $\mathcal{E}_\mathrm{me}$ for the phenomenological \ac{ME} energy density.

By comparison with Eq.~\eqref{eq:me_energy_SM}
we obtain the following relations between the \ac{ME} constants associated with anisotropy and the microscopic parameters,
\begin{align}
    B_1 &= \frac{1}{3V_c}
    \sum_\alpha
     \sum_{k\in \mathrm{N}(i)}
     \mathcal{J}_{ii,k}^{\alpha\alpha,\alpha}
    R^\alpha_{ik}\\
    B_2 &= \frac{1}{12V_c}
    \sum_{\alpha}
    \sum_{\beta, \beta\neq\alpha}
     \sum_{k\in \mathrm{N}(i)}
     \Big(
     \mathcal{J}_{ii,k}^{\alpha \beta, \alpha} R^\beta_{ik}
     +
     \mathcal{J}_{ii,k}^{\alpha \beta, \beta} R^\alpha_{ik}
     \Big).
\end{align}

The \ac{ME} constants $A_1$ and $A_2$ stem from the isotropic (Heisenberg-type) contribution of the $J_{ij,k}^{\alpha\beta,\mu}$, which can be written as $\delta^{\alpha\beta} \tilde{A}^{\mu}_{ij,k}$. Inserting this expression into Eq.~\eqref{eq:expanded_ssl_SM} and following the derivations of $B_1$ and $B_2$ given above, we calculate

\begin{align}
\mathcal{H}^\mathrm{iso}_\mathrm{ssl}
&=
\sum_{i}
\sum_{jk\in \mathrm{N}(i)}
\sum_{\alpha\mu\nu}
\tilde{A}^\mu_{ij,k}
S_i^\alpha
S_j^\alpha
R_{ik}^\nu
\partial_\nu u_i^\mu
=
E_0 -
\sum_{i}
\sum_{jk\in \mathrm{N}(i)}
\sum_{\alpha\mu\nu}
\frac{\tilde{A}^\mu_{ij,k}}{2}
(S_i^\alpha-S_j^\alpha)^2
R_{ik}^\nu
\partial_\nu u_i^\mu
\\
&\approx
E_0 -
\sum_{i}
\sum_{jk\in \mathrm{N}(i)}
\sum_{\alpha\beta\gamma\mu\nu}
\frac{\tilde{A}^\mu_{ij,k}}{2}
 R_{ij}^\beta\partial_\beta S_i^\alpha
 R_{ij}^\gamma\partial_\gamma S_i^\alpha
R_{ik}^\nu
\partial_\nu u_i^\mu
\\
&\approx
E_0 -
\frac{1}{V_c}
\int \mathrm{d}V\
\sum_{\alpha\beta\gamma\mu\nu}
\partial_\beta S^\alpha(\vec{r})
\partial_\gamma S^\alpha(\vec{r})
\partial_\nu u^\mu(\vec{r})
\sum_{jk\in \mathrm{N}(i)}
\frac{\tilde{A}^\mu_{ij,k}}{2}
 R_{ij}^\beta
 R_{ij}^\gamma
R_{ik}^\nu
\\
&=
E_0 +
\int \mathrm{d}V\ 
\sum_{\alpha\beta\gamma\mu\nu}
\partial_\beta S^\alpha(\vec{r})
\partial_\gamma S^\alpha(\vec{r})
\partial_\nu u^\mu(\vec{r})
A_{\beta\gamma,\mu\nu}
\end{align}
where $E_0$ is a magnetization-independent contribution to the elastic energy and where we approximate $S_j^\alpha \approx S_i^\alpha +\sum_\nu R_{ij}^\nu \partial_\nu S_i^\alpha$. The \ac{ME} exchange tensor is given by
\begin{equation}
   A_{\beta\gamma,\mu\nu}
   =
   -\frac{1}{2V_c}
   \sum_{jk\in \mathrm{N}(i)}
\tilde{A}^\mu_{ij,k}
 R_{ij}^\beta
 R_{ij}^\gamma
R_{ik}^\nu.
\end{equation}

Again, we separate this tensor into a symmetric and an antisymmetric part (with respect to the indices $\mu,\nu$) via
\begin{align}
    A^\mathrm{s}_{\beta\gamma,\mu\nu}
   &=
   -\frac{1}{4V_c}
   \sum_{jk\in \mathrm{N}(i)}
    R_{ij}^\beta
    R_{ij}^\gamma
    \Big(
    \tilde{A}^\mu_{ij,k}
    R_{ik}^\nu
    +
    \tilde{A}^\nu_{ij,k}
    R_{ik}^\mu
    \Big)\\
    A^\mathrm{as}_{\beta\gamma,\mu\nu}
   &=
   -\frac{1}{4V_c}
   \sum_{jk\in \mathrm{N}(i)}
    R_{ij}^\beta
    R_{ij}^\gamma
    \Big(
    \tilde{A}^\mu_{ij,k}
    R_{ik}^\nu
    -
    \tilde{A}^\nu_{ij,k}
    R_{ik}^\mu
    \Big)
\end{align}
and introduce infinitesimal strain and rotation tensors. The \ac{ME} energy density associated with exchange henceforth becomes
\begin{align}
   \epsilon^\mathrm{iso}
   =
   \sum_{\alpha\beta\gamma\mu\nu}
   \partial_\beta S^\alpha
   \partial_\gamma S^\alpha
   \Big(
   A^\mathrm{s}_{\beta\gamma,\mu\nu}
   \varepsilon_{\mu\nu}
   +
   A^\mathrm{as}_{\beta\gamma,\mu\nu}
   \omega_{\mu\nu}
   \Big).
\end{align}
By comparison with Eq.~\eqref{eq:me_energy_SM}
we obtain the following relations between the \ac{ME} constants associated with exchange and the microscopic parameters,
\begin{align}
    A_1
    &=
    -\frac{1}{24V_c}
    \sum_{\beta}
    \sum_{\mu,\mu\neq\beta}
   \sum_{jk\in \mathrm{N}(i)}
    R_{ij}^\beta
    R_{ij}^\mu
    \Big(
    \tilde{A}^\mu_{ij,k}
    R_{ik}^\beta
    +
    \tilde{A}^\beta_{ij,k}
    R_{ik}^\mu
    \Big)
    \\
    A_2 
    &=
    -\frac{1}{6 V_c}
    \sum_{\beta}
    \sum_{jk\in \mathrm{N}(i)}
    \tilde{A}^\beta_{ij,k}
    R_{ij}^\beta
    R_{ij}^\beta
    R_{ik}^\beta.
\end{align}

In Ref.~\cite{Mankovsky2022} it was demonstrated that distortions of the lattice can induce substantial \ac{DMI} in bcc Fe, even though in equilibrium this type of exchange is forbidden by symmetry. However, contributions of the \ac{DMI} to the \ac{ME} energy density -- to the best of our knowledge -- have not been considered so far. On the level of the microscopic spin lattice Hamiltonian, displacement-induced \ac{DMI} is modeled by the antisymmetric (with respect to spins) part of the $J_{ij,k}^{\alpha\beta\gamma}$. Using the Levi-Civita tensor $\varepsilon^{\alpha\beta\gamma}$ it can be expressed as $(J^{\mathrm{DMI}})_{ij,k}^{\alpha\beta,\mu}=\sum_{\gamma}\varepsilon^{\alpha\beta\gamma}D^{\gamma,\mu}_{ijk}$, where $D^{\gamma,\mu}_{ijk}$ represents the DM vector for the interaction between the spins $i$ and $j$ induced by the relative displacement of the atom $k$ in the direction $\mu$. The corresponding Hamiltonian can be brought in continuum form following the calculations above:
\begin{align}
    \mathcal{H}^{\mathrm{DMI}}_{\mathrm{ssl}}
    &=
    \sum_{ijk}\sum_{\alpha\beta\gamma\mu\nu}
    \varepsilon^{\alpha\beta\gamma}D^{\gamma,\mu}_{ij,k}
    S_i^\alpha
    S_j^\beta
    R_{ik}^\nu
    \partial_\nu
    u_i^\mu
    \approx
    \sum_{ijk}\sum_{\alpha\beta\gamma\mu\nu}
    \varepsilon^{\alpha\beta\gamma}D^{\gamma,\mu}_{ij,k}
    S_i^\alpha
    (S_i^\beta + \sum_\delta R_{ij}^\delta \partial_\delta S_i^\beta)
    R_{ik}^\nu
    \partial_\nu
    u_i^\mu
    \\
    &=
    \sum_{i}
    \sum_{\alpha\beta\gamma\delta\mu\nu}
    \varepsilon^{\alpha\beta\gamma}
    S_i^\alpha
    \partial_\delta S_i^\beta
    \partial_\nu u_i^\mu
    \sum_{jk\in\mathrm{N}(i)}
    D^{\gamma,\mu}_{ij,k}
    R_{ij}^\delta 
    R_{ik}^\nu
    \\
    &\approx
    \frac{1}{V_c}
    \int \mathrm{d}V\
    \sum_{\alpha\beta\gamma\delta\mu\nu}
    \varepsilon^{\alpha\beta\gamma}
    S^\alpha(\vec{r})
    \partial_\delta S^\beta(\vec{r})
    \partial_\nu u^\mu(\vec{r})  
    \sum_{jk\in\mathrm{N}(i)}
    D^{\gamma,\mu}_{ij,k}
    R_{ij}^\delta 
    R_{ik}^\nu
    \\
    &=
    \int \mathrm{d}V\
    \sum_{\alpha\beta\gamma\delta\mu\nu}
    \varepsilon^{\alpha\beta\gamma}
    S^\alpha(\vec{r})
    \partial_\delta S^\beta(\vec{r})
    \partial_\nu u^\mu(\vec{r}) 
    D_{\gamma\delta,\mu\nu}.
\end{align}

The \ac{ME} \ac{DMI} tensor is defined as
\begin{equation}
    D_{\gamma\delta,\mu\nu}
    =
    \frac{1}{V_c}
    \sum_{jk\in\mathrm{N}(i)}
    D^{\gamma,\mu}_{ij,k}
    R_{ij}^\delta 
    R_{ik}^\nu
\end{equation}
and its symmetric and antisymmetric parts read
\begin{align}
    D^\mathrm{s}_{\gamma\delta,\mu\nu}
    &=
    \frac{1}{2V_c}
    \sum_{jk\in\mathrm{N}(i)}
    R_{ij}^\delta 
    \Big(
    D^{\gamma,\mu}_{ij,k}
    R_{ik}^\nu
    +
    D^{\gamma \nu}_{ij,k}
    R_{ik}^\mu
    \Big)
    \\
    D^\mathrm{as}_{\gamma\delta,\mu\nu}
    &=
    \frac{1}{2V_c}
    \sum_{jk\in\mathrm{N}(i)}
    R_{ij}^\delta 
    \Big(
    D^{\gamma,\mu}_{ij,k}
    R_{ik}^\nu
    -
    D^{\gamma,\nu}_{ij,k}
    R_{ik}^\mu
    \Big).
\end{align}
Altogether, the \ac{DMI}-like contribution to the \ac{ME} energy density reads
\begin{align}
\begin{split}
    \epsilon^\mathrm{DMI}
    =
    \sum_{\alpha\beta\gamma\delta\mu\nu}
    \varepsilon^{\alpha\beta\gamma}
    S^\alpha
    \partial_\delta S^\beta
    \Big(
    D^\mathrm{s}_{\gamma\delta,\mu\nu}
    \varepsilon_{\mu\nu}
    +
    D^\mathrm{as}_{\gamma\delta,\mu\nu}
    \omega_{\mu\nu}
    \Big).
\end{split}
\end{align}
Note that the terms proportional to the strain tensor can only exist in systems with broken inversion symmetry.


The framework we have established here goes beyond the conventional formulation of magneto-elastic theory \cite{Kittel1949}. Based on the microscopic spin-lattice Hamiltonian \cite{Hellsvik2019,Mankovsky2022} we have derived general expressions describing the magneto-elastic energy density that are first order in the strain tensor as well as the rotation tensor, the importance of which has been highlighted before by Melcher \cite{Melcher1972}. Our approach also yields closed expressions for the magneto-elastic constants $B_1$, $B_2$, $A_1$ and $A_2$, which are obtained by summing over all neighbors and spatial directions. As such, their relation to the microscopic spin-lattice parameters can be viewed as analogous to the relation between the microscopic spin-spin interactions and the micromagnetic parameters \cite{Exl2020}.


\section{Simulation method}
Our combined spin-lattice dynamics simulations require the concurrent solution of the following coupled equations of motion,
\begin{align}
    \dot{\vec{r}}_i = \frac{\partial \mathcal{H}}{\partial \vec{p}_i}=\frac{\vec{p}_i}{m}, \hspace{2em}
    \dot{\vec{p}}_i = -\frac{\partial \mathcal{H}}{\partial \vec{r_i}}=\vec{F}_i, \hspace{2em}
    \dot{\vec{S}}_i = -\frac{\gamma}{\mu_\mathrm{s}} \vec{S}_i \times \vec{H}^\mathrm{eff}_i,
\end{align}

where the effective field is given by $\vec{H}^\mathrm{eff}_i =-\frac{\partial \mathcal{H}}{\partial \vec{S}_i}$. To integrate these equations of motion we use a scheme based on the Liouville formalism \cite{Frenkel2001} and the Suzuki-Trotter decomposition \cite{Suzuki1976} that was initially proposed in Ref.~\cite{Omelyan2001}. This scheme can integrate non-commuting operators and conserves energy, momentum and angular momentum. 

By introducing a generalized phase-space variable $\vec{X}=\{ \vec{r}_i,\vec{p}_i,\vec{S}_i \}$, the equations of motion can be reformulated using Liouville operators \cite{Frenkel2001},

\begin{equation}
    \frac{\partial \vec{X}}{\partial t} 
    = \hat{L}\vec{X}
    = (\hat{L}_{{r}} + \hat{L}_{{p}} + \hat{L}_{{S}})\vec{X}.
    \label{eq:Liouville_equation}
\end{equation}
Formally, the solution of the Liouville equation can be written in the form $\vec{X}(t + \Delta t) = \mathrm{e}^{\hat{L} \Delta t} \vec{X}(t) = \mathrm{e}^{(\hat{L}_{{r}} + \hat{L}_{{p}} + \hat{L}_{{S}}) \Delta t} \vec{X}(t)$. As the exponential propagator cannot be evaluated exactly, one makes use of the Suzuki-Trotter decomposition \cite{Suzuki1976}, which states that 
\begin{equation}
    \mathrm{e}^{\hat{A}+\hat{B}} = \lim_{n\rightarrow\infty}(\mathrm{e}^{\frac{\hat{A}}{2n}} \mathrm{e}^{\frac{\hat{B}}{n}} \mathrm{e}^{\frac{\hat{A}}{2n}})^n,
\end{equation}
with $\hat{A}$ and $\hat{B}$ being two non-commuting operators. For finite $n$, the above expression becomes
\begin{equation}
    \mathrm{e}^{\hat{A}+\hat{B}} = (\mathrm{e}^{\frac{\hat{A}}{2n}} \mathrm{e}^{\frac{\hat{B}}{n}} \mathrm{e}^{\frac{\hat{A}}{2n}})^n \mathrm{e}^{\mathcal{O}(n^{-2})},
\end{equation}
This decomposition can be abbreviated as $(A,B,A)$. Note that the Suzuki-Trotter decomposition is not unique, since the decomposition $(B,A,B)$ is also possible. 

Coming back to the Liouville equation for the spin-lattice dynamics~\eqref{eq:Liouville_equation}, it is apparent that there are multiple alternatives for the decomposition of the exponetial propagator.
Omelyan et al.~\cite{Omelyan2001} proposed the decomposition $(\vec{p},\vec{r},\vec{S},\vec{r},\vec{p})$, whereas Tsai et al.~\cite{Tsai2005} used $(\vec{S},\vec{p},\vec{r},\vec{p},\vec{S})$. Here, we use the decomposition $(\vec{r},\vec{p},\vec{S},\vec{p},\vec{r})$,
\begin{equation}
    \vec{X}(t + \Delta t) 
    =
    \mathrm{e}^{\hat{L}_{r} \frac{\Delta t}{2}}
    \mathrm{e}^{\hat{L}_{p} \frac{\Delta t}{2}}
    \mathrm{e}^{\hat{L}_{S} \Delta t}
    \mathrm{e}^{\hat{L}_{p} \frac{\Delta t}{2}}
    \mathrm{e}^{\hat{L}_{r} \frac{\Delta t}{2}}
    \vec{X}(t)
    +
    \mathcal{O}(\Delta t^3),
\end{equation}
to minimize the frequency of evaluation of $\hat{L}_{S}$, which is the most time-consuming step of the algorithm.

Momenta and positions are updated using a first oder update, whereas the spin is updated using a Cayley transformation \cite{Lewis2003}, in order to conserve the norm of each individual spin:
\begin{align}
    \mathrm{e}^{\hat{L}_{r} \Delta t}\vec{r}_i&=\vec{r}_i + \Delta t \frac{\vec{p}_i}{m_i} , \hspace{4em}
    \mathrm{e}^{\hat{L}_{p} \Delta t}\vec{p}_i=\vec{p}_i + \Delta t \vec{F}_i \\
    \mathrm{e}^{\hat{L}_{S} \Delta t}\vec{S}_i&=
    \frac{
    \vec{S}_i + \Delta t \frac{\gamma}{\mu_\mathrm{s}} \vec{H}^\mathrm{eff}_i \times \vec{S}_i
    }{
    1 + \frac{1}{4}\Big(\Delta t\frac{ \gamma}{\mu_\mathrm{s}}\Big)^2
    (\vec{H}^\mathrm{eff}_i)^2
    }
    +\frac{
    \frac{1}{2}
    \Big(\Delta t\frac{ \gamma}{\mu_\mathrm{s}}\Big)^2
    \Big(
    (\vec{H}^\mathrm{eff}_i \cdot \vec{S}_i)
    \vec{H}^\mathrm{eff}_i
    -\frac{1}{2}
    (\vec{H}^\mathrm{eff}_i)^2 \vec{S}_i
    \Big)}{
    1 + \frac{1}{4}\Big(\Delta t\frac{ \gamma}{\mu_\mathrm{s}}\Big)^2
    (\vec{H}^\mathrm{eff}_i)^2
    }.
    \label{eq:Cayley_transform}
\end{align}
Within the Heisenberg model, the effective field $\vec{H}^\mathrm{eff}_i$ acting on the spin at site $i$ depends on the neighboring spin orientations and henceforth an additional Suzuki-Trotter decomposition of the form 
\begin{equation}
    \mathrm{e}^{\hat{L}_{S} \Delta t}
    =
    \mathrm{e}^{\hat{L}_{S_1} \frac{\Delta t}{2}}
    \mathrm{e}^{\hat{L}_{S_2} \frac{\Delta t}{2}}
    \dots
    \mathrm{e}^{\hat{L}_{S_N} \Delta t}
    \dots    
    \mathrm{e}^{\hat{L}_{S_2} \frac{\Delta t}{2}}
    \mathrm{e}^{\hat{L}_{S_1}\frac{\Delta t}{2}}
\end{equation}
is needed \cite{Omelyan2001}. If also on-site anisotropy terms are included, the effective field also depends on the spin itself. This requires an additional, third Suzuki-Trotter decomposition of the propagator of the spin at site $i$ into
\begin{equation}
    \mathrm{e}^{\hat{L}_{S_i} \frac{\Delta t}{2}}
    =
    \mathrm{e}^{\hat{L}^{\mathrm{ani},1}_{S_i} \frac{\Delta t}{4}}
    \dots
    \mathrm{e}^{\hat{L}^{\mathrm{ani},M}_{S_i} \frac{\Delta t}{4}}
    \mathrm{e}^{\hat{L}^{\mathrm{Hei}}_{S_i} \frac{\Delta t}{2}}
    \mathrm{e}^{\hat{L}^{\mathrm{ani},M}_{S_i} \frac{\Delta t}{4}}
    \dots
    \mathrm{e}^{\hat{L}^{\mathrm{ani},1}_{S_i} \frac{\Delta t}{4}}.
\end{equation}
Here, $\mathrm{e}^{\hat{L}^{\mathrm{ani},k}_{S_i}}$ with $k\in \{1,M\}$ describes the propagation of the spin at lattice site $i$ due the effective field generated by the $k$-th contribution to the on-site anisotropy (for the Hamiltonian used in our simulations, Eqs.~(5) and (9) in the main text, we have $M=2$) via the Cayley transformation~\eqref{eq:Cayley_transform}. The term $\mathrm{e}^{\hat{L}^{\mathrm{Hei}}_{S_i}}$ describes the propagation in the combined effective field generated by all neighbors. For this term, an additional Suzuki-Trotter decomposition is not needed.

In summary, our three-level Suzuki-Trotter-decomposed integration scheme can be abbreviated as
\begin{align}
    \begin{split}
        &\hspace*{19.ex}(\vec{r},\vec{p},\vec{S},\vec{p},\vec{r})\\
        &\hspace*{7.25ex}\overbrace{(\vec{S}_1,\dots,\vec{S}_i,\dots, \vec{S}_N,\dots,\vec{S}_i,\dots,\vec{S}_1)} \\
        &\overbrace{\mathrm{ani}_1,\dots,\mathrm{ani}_M,\mathrm{Hei},\mathrm{ani}_M,\dots,\mathrm{ani}_1}.
    \end{split}
\end{align}
Tests of the accuracy of the integration have been performed by checking the conservation of the total angular momentum and the energy of the system (see below).

\section{Numerical tests}
Hereinafter we test the accuracy of our integration scheme for the simulation of combined spin-lattice dynamics for the model Hamiltonian introduced in the main text,
\begin{align}
    \begin{split}
        \mathcal{H}=&-\sum_{ij} J_{ij} \vec{S}_i \cdot  \vec{S}_j  
        + \sum_{i} \frac{\vec{p}_i^2}{2 m}
        + V_0 \sum_{ij} \frac{(r_{ij}-R_{ij})^2}{R_{ij}} 
        -\frac{d_z}{2}\sum_i \Big[\Big(\vec{S}_i\cdot \frac{\vec{r}_i^{z+}-\vec{r}_i}{\vert\vec{r}_i^{z+}-\vec{r}_i \vert }\Big)^2
     +\Big(\vec{S}_i\cdot \frac{\vec{r}_i^{z-}-\vec{r}_i}{\vert\vec{r}_i^{z-}-\vec{r}_i \vert }\Big)^2 \Big].
    \end{split}
    \label{eq:Hamiltonian_simulations_SM}
\end{align}
Specifically, we demonstrate the conservation of the total energy and angular momentum and show that the equilibrium magnetization obtained by spin-lattice dynamics agrees with what is obtained by conventional spin dynamics simulations. 

For our simulations we use the following set of parameters, $m=100\mu_\mathrm{S}^2/(a^2 \gamma^2 J)$, $V_0= 100 J/a$, $d_z=0.1J$, $\Delta t=0.01 \frac{\mu_\mathrm{s}}{\gamma J}$, where $a=R_{ij}^{\mathrm{n.n.}}$ is the nearest neighbor equilibrium distance, i.e.~ the lattice constant of the simple cubic system. 

\begin{figure}
    \centering
        \includegraphics[width=0.47\textwidth]{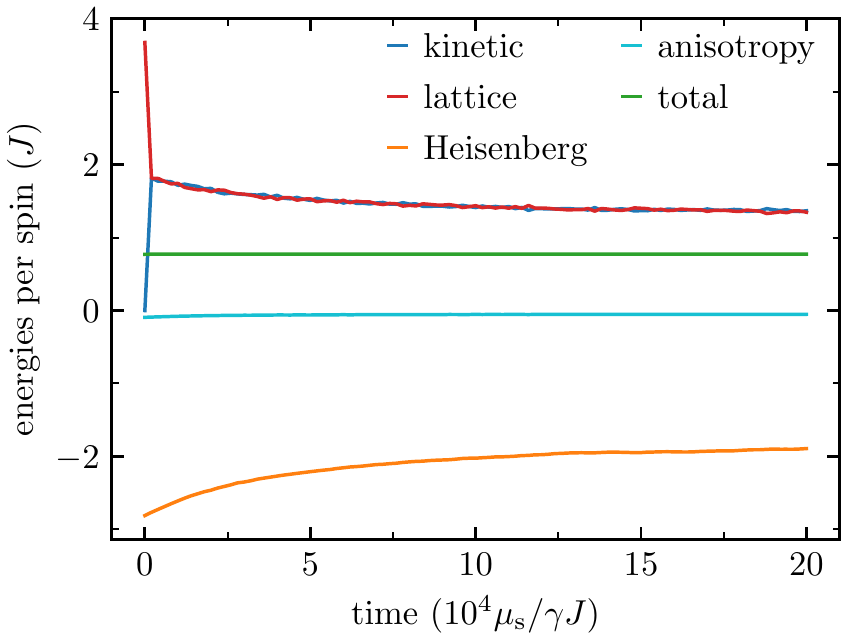}~a)   \includegraphics[width=0.47\textwidth]{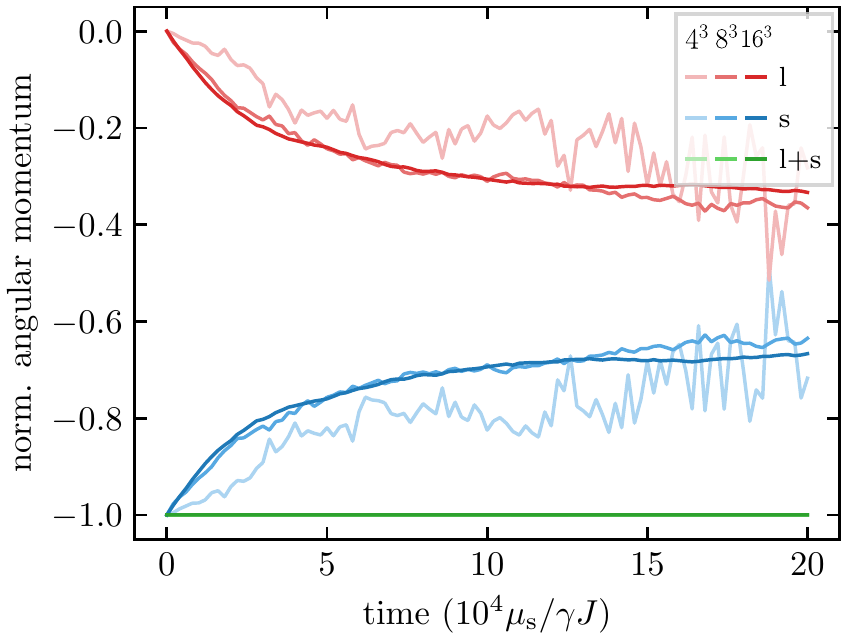}~b)

    \caption[]{Energy and angular momentum transfer between lattice and spins. a) Individual contributions to the energy versus time for cubic nanoparticle consisting of $16^3$ spins. b) Components of the angular momentum of spins and lattice along the $z$-axis versus time for different cube sizes as labeled.}
    \label{fig:energy_ang_moment_transfer}
\end{figure}

Fig.~\ref{fig:energy_ang_moment_transfer} displays the time evolution of the individual contributions to the energy and the angular momenta for a cubic nanoparticle. Initially, the spins are aligned collinearly along the easy axis and the atoms are at rest ($\vec{p}_i=0$), but randomly displaced from their equilibrium positions by up to $8 \% $ of the lattice constant. Lattice energy and kinetic energy of the atoms quickly equilibrate to the same value, in agreement with the equipartition theorem. The energy of the lattice degrees of freedom decrease gradually, accompanied with an increase in the Heisenberg energy. Note that, however, the total energy stays constant.

If the initial spin configuration is collinear, this increase in Heisenberg exchange is inevitably associated with a change in the magnetization. This is displayed in Fig.~\ref{fig:energy_ang_moment_transfer}b for different sizes of the nanoparticle. Physically speaking, the energy transfer from the lattice to the spins heats up the latter, giving rise to demagnetization. Since angular momentum is conserved, this is associated with a transfer of angular momentum from the spins to the lattice, i.e.\ the lattice will eventually start to rotate around the initial orientation of the magnetization, following the Einstein-de-Haas effect \cite{Einstein1915}.

\begin{figure}
    \centering
    \includegraphics[width=0.48\textwidth]{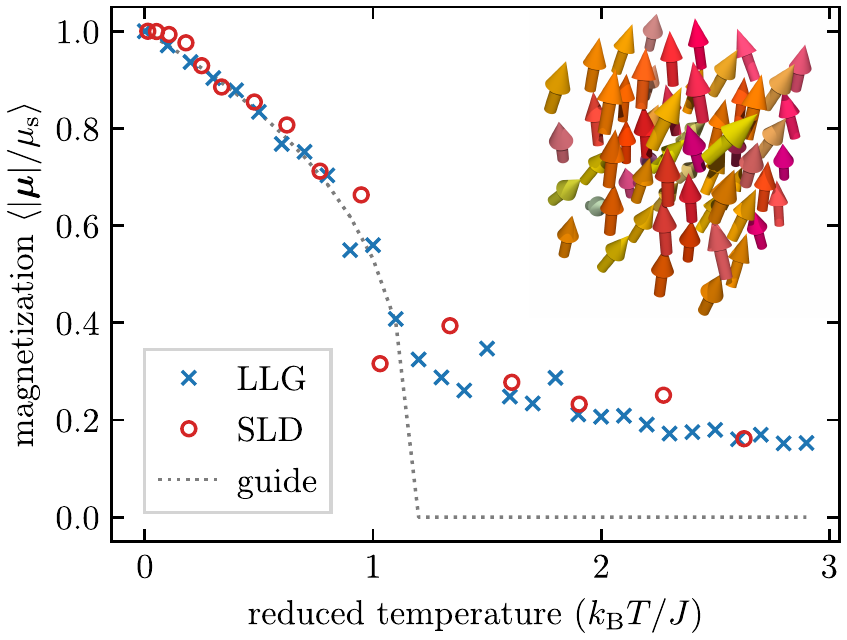}
    \caption{Magnetization versus temperature for a cubic nanoparticle consisting of $4^3$ spins obtained from spin dynamics simulations using the stochastic Landau-Lifshitz-Gilbert equation (LLG) and spin lattice dynamics (SLD). The dotted line corresponds to the finite-size scaled prediction for a three-dimensional Heisenberg model.}
    \label{fig:m_vs_T}
\end{figure}

As a second test we investigate the temperature-dependence of the equilibrium magnetization. This can be achieved by using the same simulation setup as above for different values of the initial random lattice displacement. Both the temperature and the magnetization are obtained by averaging for $2 \times 10^5 \frac{\mu_\mathrm{s}}{\gamma J}$ after an initial equilibration for $2 \times 10^5 \frac{\mu_\mathrm{s}}{\gamma J}$. The temperature is calculated from the kinetic energy of the lattice using the relation $\langle E_\mathrm{kin}\rangle=\frac{3}{2}k_\mathrm{B}T$. In addition, we perform spin dynamics simulations based on the stochastic Landau-Lifshitz Gilbert equation 
\begin{equation}
    \frac{\partial\vec{S}_i}{\partial t} = 
    -\frac{\gamma}{(1 + \alpha^2)\mu_\mathrm{s}} 
    \vec{S}_i \times 
    \Big(
    (\vec{H}^\mathrm{eff}_i + \vec{\zeta}_i)
    +
    \alpha 
    \vec{S}_i
    \times
    (\vec{H}^\mathrm{eff}_i + \vec{\zeta}_i)
    \Big),
\end{equation}
with $\alpha$ being the Gilbert damping parameter and where $\vec{\zeta}_i$ is a white noise term with the properties $\langle \vec{\zeta}_i\rangle=0$ and $\langle \zeta^\mu_i(t)\zeta^\nu_j(0)\rangle = 2 \alpha k_\mathrm{B} T\frac{\mu_\mathrm{s}}{\gamma}  \delta_{ij}\delta_{\mu\nu}\delta(t)$. The results are summarized in Fig.~\ref{fig:m_vs_T}. We find that the equilibrium magnetization is the same for the pure spin model calculations and the spin-lattice dynamics. The dotted line corresponds to the phenomenological relation $\langle |\vec{\mu}|/\mu_\mathrm{s}\rangle=(1-T/T_\mathrm{C}(l))^{\frac{1}{3}}$ for three-dimensional Heisenberg models with the finite-size scaled Curie temperature $T_\mathrm{C}(l)$ \cite{Ellis2015}. For a cubic system consisting of $4^3$ spins it can be calculated that $k_\mathrm{B}T_\mathrm{C}(l) = 1.173 J$ \cite{Rozsa2019}.

\section{Ab initio Calculations \label{appendix:SPRKKR}}
The results for spin-spin and spin-lattice exchange coupling tensors presented in Table I of the main text are based on first-principles electronic structure calculations using the spin-polarized relativistic Korringa Kohn Rostoker (SPRKKR) Green function method \cite{SPR-KKR8.5,Ebert2011}. 
The elements of the SSC tensor were calculated using the relativistic extension of the Lichtenstein formula \cite{Liechtenstein1987} as presented in Refs. \cite{Ebert2009, Udvardi2003} and are given by the expression 
\begin{eqnarray}
 J_{ij}^{\alpha \beta}  &=&  \frac{1}{2\pi} \mathrm{Im}\, \mathrm{Tr} \int dE\, 
\underline{T}^{\alpha}_i \underline{\tau}^{ij}
\underline{T}^{\beta}_j \underline{\tau}^{ji} \; 
\label{eq:Jij_stand}
\end{eqnarray}
with $\underline{\tau}^{ij}$ the scattering path operator connecting  sites $i$ and $j$
with the underline indicating matrices in the $\Lambda=(\kappa,\mu)$-representation \cite{Rose1961}, and the
corresponding on-site coupling for sites $i$ and $j$ represented by the matrix
\begin{eqnarray}
T^{i\alpha}_{\Lambda\Lambda'} & = & \int d^3r
Z^{\times}_{i,\Lambda}(\vec{r})\beta \sigma_{\alpha}B_i(r)
Z_{i,\Lambda'}(\vec{r}) \; .
\label{eq:Jij_ME}
\end{eqnarray}
Here $\beta$ is one of the standard 
Dirac matrices, $\sigma_{\alpha}$ is a $4 \times 4$-Pauli matrix \cite{Rose1961}
and $B(r)$ is the spin-dependent part of the exchange-correlation
potential set up within local spin-density theory, and
$Z_{\Lambda}(\vec{r})$ are the solutions
of the Dirac equation normalized according to the
relativistic multiple-scattering formalism \cite{Ebert2011}.

The SLC tensors are calculated using the closed expressions for the SLC tensor elements that were recently worked out by Mankovsky et al. \cite{Mankovsky2022} and given by
\begin{eqnarray}
  {\cal J}^{\alpha\beta,\mu}_{ij,k} &=& \frac{\partial^3 {\cal F}}{\partial e^\alpha_i \,\partial e^\beta_j
                                        \, \partial u^{\mu}_k} = \frac{1}{2\pi} \mbox{Im\, Tr\,}
                                  \int^{E_F}dE \,\, 
                           \Big[   \underline{T}^{\alpha}_i \,\underline{\tau}_{ij}
                                  \underline{T}^{\beta}_j 
                                 \, \underline{\tau}_{jk}
                                \underline{\cal U}^{\mu}_k \,\underline{\tau}_{ki}
                             +  \underline{T}^{\alpha}_i
                                   \,\underline{\tau}_{ik}
       \underline{\cal U}^{\mu}_k \,\underline{\tau}_{kj}
                                  \underline{T}^{\beta}_j 
                                 \, \underline{\tau}_{ji}\Big] \,.\quad\quad\,
       \label{eq:Parametes_linear} 
\end{eqnarray}
The derivation of this SLC formula is based on calculating the energy change in the presence of spin tiltings $\delta \hat{e}_i^\mu$ and
atomic displacements of the atoms ${u}_i^\nu$ and considering the related change of the scattering path operator in a perturbative manner. The resulting change of the inverse $t$-matrix, denoted by $\underline{\underline{m}}$, is written as
\begin{eqnarray}
  \Delta^s_{\alpha} {\underline{m}}_i  =   \delta \hat{e}^\alpha_i\,
       \underline{T}^{\alpha}_i \,, \hspace{2em}
  \Delta^{u}_{\mu} \underline{m}_i  =  
  u^\mu_i  \underline{\cal U}_{i}^{\mu}  \,,
\label{eq:linear_distor}
\end{eqnarray}
where the auxiliary matrices $\underline{T}^{\alpha}_i $ and ${\cal
  U}_{i}^{\nu}$ \cite{Papanikolaou97} have been introduced.
The latter one is given by
\begin{eqnarray}
\underline{\cal U}_{k}^{\nu} =\bar{\underline{U}}(\hat{u}^{\,\nu}_k)\,\underline{m}_k+\underline{m}_k\,\bar{\underline{U}}(-\hat{u}^{\,\nu}_k)
\end{eqnarray}
with the transformation matrices
\begin{eqnarray}
U_{LL'}(\vec{u}_k)& = &
\delta_{LL'} + 
|\vec{u}_k|\bar{U}_{LL'}(\hat{u}_k)\,,
\end{eqnarray}
where
\begin{eqnarray*}
%
\bar{U}_{LL'}(\hat{u}_k)& = & \kappa \frac{4\pi}{3}i^{l+1-l'} 
\sum_{m=-1}^1 C_{LL'1m}  \,  Y_{1m}(\hat{u}_k) 
\end{eqnarray*}
and $\kappa = \sqrt{2mE/\hbar^2}$.\\

For all calculations, the atomic sphere approximation (ASA) as well as the local spin density approximation (LSDA) have been used with the parametrization of the exchange and correlation potential by Vosko et al. \cite{Volko80}. The self consistency loops for converging the potentials and the calculation of SSC and SLC tensors were performed with an angular momentum expansion of up to the cutoff $l_{\mathrm{max}}=3$. For bcc Fe a k-mesh with $45\times 45\times 45$ grid points and for FePt with $35\times 35\times 24$ grid points were used for the integration over the Brillouin zone.


%